# Detection of the melanoma biomarker TROY using silicon nanowire field-effect transistors

*Carsten Maedler,[1] Daniel Kim,[1] Remco A. Spanjaard,[2] Mi Hong,[1] Shyamsunder Erramilli,[1,3] Pritiraj Mohanty[1,]\**

[1]Department of Physics, Boston University, 590 Commonwealth Avenue, Boston, MA 02215, USA, [2]Femto Diagnostics, 53 Bay State Road, Boston, MA 02215, [3]Department of Biomedical Engineering & Photonics Center, Boston University, 8 St. Mary's St., Boston, MA 02215, USA

*Keywords:* Nanowire, melanoma, TROY, biomarker testing, biosensor, molecular gating

**Abstract:** Antibody-functionalized silicon nanowire field-effect transistors have been shown to exhibit excellent analyte detection sensitivity enabling sensing of analyte concentrations at levels not readily accessible by other methods. One example where accurate measurement of small concentrations is necessary is detection of serum biomarkers, such as the recently discovered tumor necrosis factor receptor superfamily member TROY (TNFRSF19), which may serve as a biomarker for melanoma. TROY is normally only present in brain but it is aberrantly expressed in primary and metastatic melanoma cells and shed into the surrounding environment. In this study, we show the detection of different concentrations of TROY in buffer solution using top-down fabricated silicon nanowires. We demonstrate the selectivity of our sensors by comparing the signal with that obtained from bovine serum albumin in buffer solution. Both the signal size and the reaction kinetics serve to distinguish the two signals. Using a fast-mixing two-compartment reaction model, we are able to extract the association and dissociation rate constants for the reaction of TROY with the antibody immobilized on the sensor surface.

## I. Introduction

While the incidence of many common cancers is decreasing, according to the American Cancer Society the incidence of melanoma has been rising by about 3% per year. Statistical analysis shows there will be an estimated 76,690 new cases in 2013, with 9,480 deaths. If melanoma is detected early, the five-year survival rate is 98% but this rate rapidly decreases to 62% for regional disease and to 15% and lower when tumors have spread to distant sites. While treatment options for advanced stage melanoma had stagnated for decades, recently a number of new targeted therapies have become available, but, unfortunately, even these drugs are only expected to increase survival by several months in certain patient groups before they too inevitably relapse.[1,2] If an early diagnosis can be made before metastatic disease becomes clinically detectable and virtually incurable, it may be possible to effectively treat patients because the tumor burden is still very low. But presently no such early detection method exists. Recently, it was found that a member of the tumor necrosis factor receptor superfamily, called TROY (or TNFRSF19), may serve as a melanoma-specific biomarker.[3] TROY is widely expressed during embryogenesis, but in human adults it is primarily restricted to the brain. Studies have shown that TROY is expressed in all investigated primary and metastatic melanoma cells and tissue samples, but not in melanocytes found in normal skin biopsies and primary skin cell cultures. Neither is it detectable in other (skin) tumor cells.[3] Many membrane receptors are shed into the surrounding environment[4,5] and our preliminary studies showed that the extracellular domain (ECD) of TROY is also shed (data not shown). This suggests that TROY may serve as a novel surrogate serum biomarker for circulating and metastatic melanoma cells and a Point-Of-Care (POC) diagnostic blood test that can detect abnormally elevated levels of TROY ECD may help revolutionize early diagnosis and treatment for this disease. This project intends to address this huge unmet clinical need by first generating and characterizing a nanosensor dedicated to detection of low quantities of TROY in vitro as a first step towards developing this sensor technology towards clinical use as a novel POC device to detect occult metastatic disease.

With recent advances in nanotechnology many novel biosensing systems have been developed.[6] Among those different techniques for biomarker sensing especially electrical detection using ion-sensitive Field-Effect Transistors (FET) holds a great promise to revolutionize biosensing[7] due to its direct, label-free[8] sensing and highly scalable nature.[9] Silicon nanowire FETs have been used to detect pH,[10,11,12,13] various cancer biomarkers,[14,15,16,17] biomarkers indicative of acute myocardial infarction,[18,19,20,21] glucose[22] and viruses.[23,24] Sensitivity of detection of molecules is enhanced by their high surface-to-volume ratio, permitting nanowires to sense surface binding events hard to achieve by bulk detection schemes. Specificity is generally achieved by immobilization of antibodies to the nanowire surface by silanization of the sample and subsequent covalent binding of the antibody through amide bonds. The binding of antigens to the paratope of the immobilized antibodies changes the charge distribution in the biofunctionalized layer. This can either happen because the antigen is charged in solution or because of the adsorption process. The result is a change of the electric field through the nanowire, which leads to accumulation or depletion of charge carriers, hence modulating its conductance.[25] For obvious reasons, this is also termed molecular gating.[26] The direct sensing of molecules of interest eliminates the need for labeling and long incubation times reducing the cost and time needed for detection. The data is essentially recorded in real time, which allows the evaluation of reaction kinetics. This should also enable distinguishing between specific and nonspecific binding due to their different reaction kinetics. The nanowires' ultra-low power consumption makes them ideal candidates for implantable biosensors to control biological function *in vivo*. There are several challenges associated with the use of semiconducting nanowires for the use of biosensors.[27] The major problem is electrostatic screening of charges in the biofunctionalized layer in physiological solutions due to their high electrolyte concentration. An indication of the screening range is the Debye length. It is the characteristic decay length of the potential in the Debye-Hückel approximation.[28] For physiological solutions it is less than 1 nm. Several groups have made efforts to increase this Debye length. Stern et al. used a microfluidic purification chip to pre-clean blood and transfer the antigens of interest to a purified buffer with low salt concentration.[29] Another approach is the use of high frequency to disrupt the electrical double layer.[30] A more simple solution is the use of antibody fragments to allow for binding of the biomarker of interest in the vicinity of the nanowire.[19]

In this study, we have detected the novel melanoma biomarker TROY using silicon nanowire FETs functionalized with the antibody fragments that are antigen binding (FAB). These fragments are only a fraction of the size of the immunoglobulin, yet retain the ability to bind antigens with high binding strength. Therefore, the antigens bind in closer vicinity to the sensing element, the nanowire. The detection limit in pure low-salt buffer was less than 10 ng/ml. By real-time data recording we extracted the reaction kinetics for binding of TROY to its corresponding antibody. Nonspecific binding tests with bovine serum albumin (BSA) showed adsorption at longer time scales, allowing for further discrimination against any non-specific background signals.

## II. Experimental Approach

We have fabricated silicon nanowires in a top-down approach starting with SOI wafer (100 nm top silicon layer, 10 Ωcm, B-doped, from Soitec) which is diced and cleaned with acetone, isopropyl alcohol and a 2:3 mixture of hydrogen peroxide and sulfuric acid (piranha, all from J.T. Baker). First, gold electrodes are formed by photolithography (Karl Suss, MA 6) and electron beam evaporation. A small layer of titanium (10 nm) is used for adhesion of the gold layer (50 nm) and a chromium layer is evaporated last as mask for subsequent etching (all from Kurt J. Lesker). Then, the nanowires (usually a set of 20 single wires about 300 nm in width and 100 nm in height) are defined by electron beam lithography (JEOL JSM 6400) and a chromium layer is evaporated as mask for isotropic dry etching in a reactive ion etcher (Plasma-Therm 790) using tetrafluoromethane and oxygen. Afterwards, the chromium layer is etched off chemically (chromium etchant 1020 from Transene), which leaves silicon nanowires on silicon oxide electrically connected by gold electrodes. In order to use the device in solution, the electrodes and wires need to be electrically insulated. This is achieved by growing 100 nm of aluminum oxide on the electrodes and 20 nm of aluminum oxide on the wires using atomic layer deposition (CambridgeNanoTech, Savannah 100). Additionally, a micrometer thick layer of poly(methyl methacrylate) (PMMA, from MicroChem) is overexposed on top of the electrodes by e-beam lithography. A finished set of wires is shown in Figure 1a. The nanowires gain selectivity by immobilized antibodies on the nanowire surface. For that purpose the surface is hydroxylated with an oxygen plasma in a plasma asher (PVA TePla, M4L, 300 mW power, 300 sccm flow rate for 2 minutes) and silanized by immersion in ethanol containing 5% (3-aminopropyl)triethoxysilane (APTES, both from Sigma-Aldrich) and 5% water for 20 minutes followed by 30 min annealing at 110 °C. The antibody FABs (provided by Biosite Diagnostics) were generated by proprietary phage-display against the TROY ECD. They are N-hydroxysuccinimide (NHS)-activated by adding them to an MES (from Sigma-Aldrich) buffer containing NHS (from Pierce) and 1-ethyl-3-(3-dimethylaminopropyl)carbodiimide (EDC, from Sigma-Aldrich). They are then reacted with the amine groups on the surface of the nanowires for 2 hours (Figure 1b). Different concentrations of TROY are detected by sealing the functionalized device in a fluid chamber[31] together with an Ag/AgCl reference electrode (A-M Systems) and flowing buffer solutions with TROY over the wires (Figure 1c). The conductance through the nanowire changes upon binding of the antigen due to a change in the charge distribution. This is measured by adding a small, low-frequency AC voltage to a DC voltage using a noninverting summing circuit[31] and measuring the response using a lock-in-amplifier (EG&G 7260) referenced to the same frequency as the AC voltage. The signal is essentially the slope of the source-drain current-voltage curve, or differential conductance.

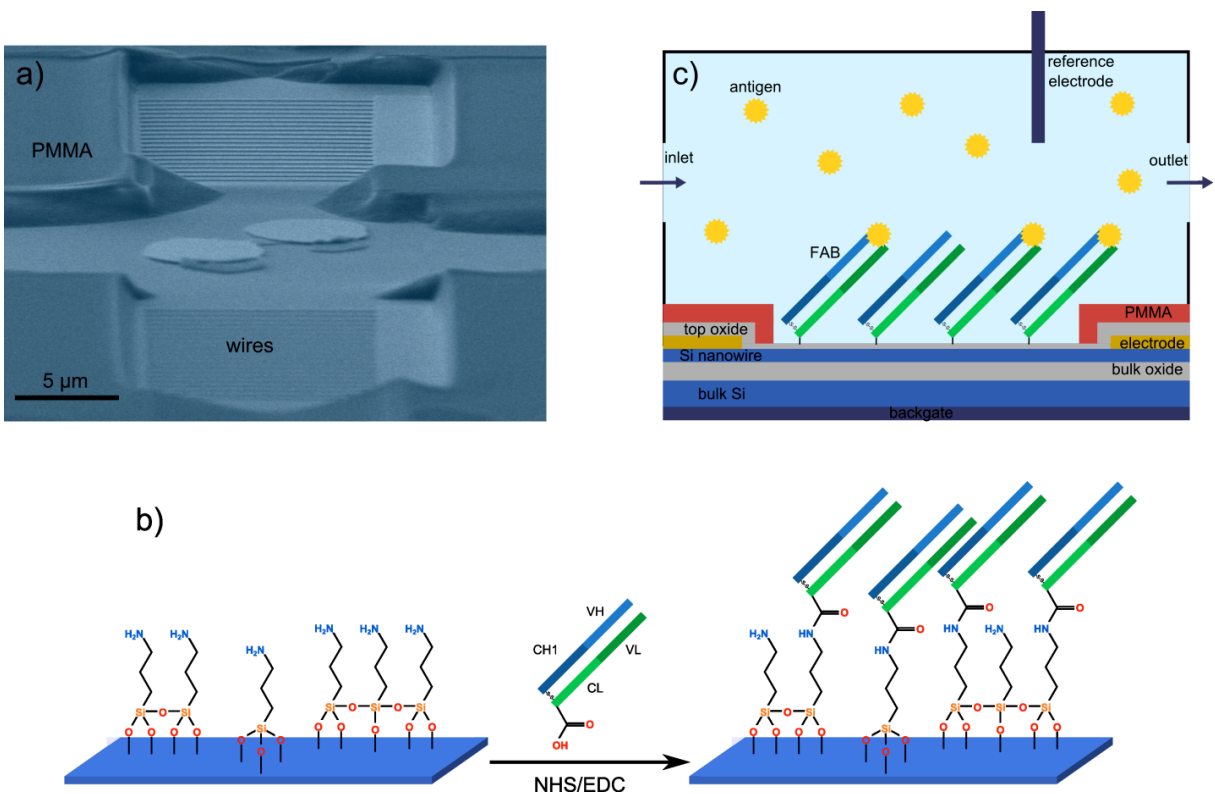

**Figure 1.** (a) Electron micrograph of two separate sets of silicon nanowires with the electrical connections insulated from solution by a micrometer thick layer of poly(methyl methacrylate) (PMMA). (b) Immobilization of the antibody FAB to the silanized nanowire using NHS/EDC coupling chemistry. The FAB has one constant and one variable domain from each heavy and light chain of the antibody (constant heavy (CH), constant light (CL), variable heavy (VH) and variable light (VL). It binds with its C-terminal end to the nanowire exposing the paratope at its variable end to the solution. (c) The sensing nanowires are sealed in a fluid chamber with an inlet and outlet at opposing ends. Their conductance is tunable by a reference electrode immersed in the solution and a back gate connected to the bulk silicon.

**III. Results and Discussion**

**A. Detection of TROY**
The recently-found potential biomarker for melanoma TROY opens new opportunities for early and confident diagnosis of melanoma as well as for monitoring recurrence and spreading of the cancer and responsiveness to therapy in established patients that are undergoing treatment. In order to make full use of the biomarker a reliable, fast and cheap biosensor needs to be developed for its detection and validation. We have used silicon nanowire FETs functionalized with the antibody FABs to detect TROY in pure 2 mM PBS buffer. For that purpose buffer solutions were prepared containing different concentrations of the antigen but equal salt concentrations and pH values to ensure that the signal is due to binding of the antigen. Devices were prescreened with respect to their I-V characteristics and an optimal working point was determined.[15] Throughout one measurement the differential conductance through the sensor was recorded in real-time. After establishing a conductance baseline with pure buffer, solutions containing the antigen and a washing buffer were then injected successively into the fluid chamber sealing the device. A syringe pump was used to allow for continuous flow. For analysis of the data the change in conductance after injection of buffer containing the antigen was calculated after saturation with respect to the baseline.

The change in conductance for three different measurements with respect to TROY concentration is shown in Figure 2. The relationship is linear on a log-log plot, which means that the intersection is zero, hence there is no significant systematic error. The error is determined by the standard deviation of each signal after saturation. This also sets a limit for detection and is illustrated in Figure 2 by the patterned area. Where the line crosses into this region determines the detection limit of the device. It corresponds to less than 10 ng/ml or 200 pM. In order to test specificity pure buffer solutions with different concentrations of BSA were flown over the device. For the same concentrations that were measured with TROY the signal for the BSA samples was below the detection limit of a conductance change of 0.11 nS. However, for a BSA concentration of 0.1 mg/ml the signal size was similar to the one generated by a TROY concentration of 50 ng/ml. As will be shown later though, the signals differ significantly due to their different adsorption times.

The signal for TROY is in the linear regime of a Langmuir adsorption isotherm, which makes extraction of the equilibrium constant highly inaccurate. Assuming a linear increase of conductance with charge distribution change in the biofunctionalized layer the change in conductance is proportional to the concentration of bound antigen-antibody complex, which can be modeled by a Langmuir adsorption isotherm. Therefore, the change in conductance $\Delta G$ is proportional to

$$\Delta G \propto \frac{C_{TROY}}{C_{TROY} + K_{eq}} , \qquad (1)$$

where $C_{TROY}$ is the antigen concentration in the buffer and $K_{eq}$ is the equilibrium constant. Approximation of the curve in Figure 2 yields an equilibrium constant of $12 \pm 4$ nM.

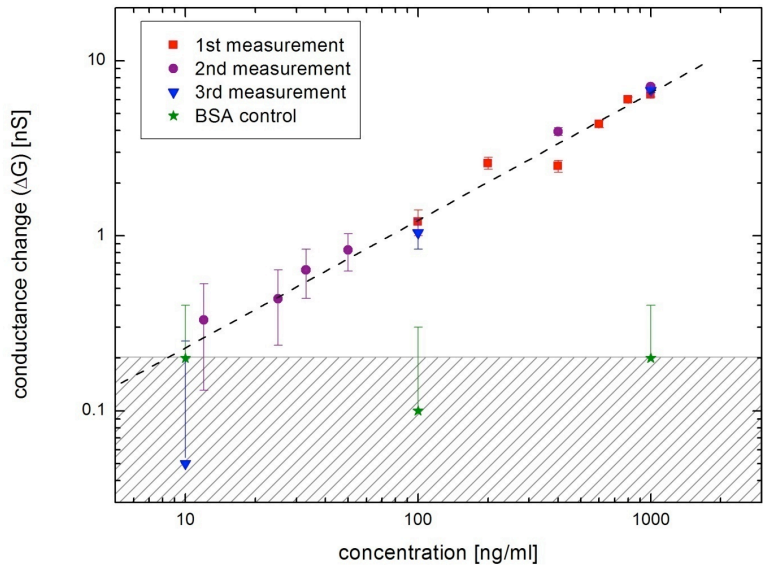

**Figure 2.** Sensing of the melanoma biomarker TROY using FAB-functionalized nanowires. The conductance change of the nanowires in response to different concentrations of TROY in 2 mM PBS buffer (squares, circles and triangles) is compared to the response to BSA (stars) in the same buffer. A linear dependence between conductance change and concentration is observed in the 10 ng/ml to 1 μg/ml regime of TROY concentration (the dashed line is a guide to the eye). The error bars are the standard deviation of the signal (they are expanded for lower concentrations due to the nature of the logarithmic plot). The largest standard deviation is 0.2 nS, which is illustrated by the shaded area. Hence, where the line crosses into the shaded area determines the detection limit. It corresponds to a concentration below 10 ng/ml.

## B. Reaction kinetics analysis

The collection of real-time data using the silicon nanowire FETs allows evaluation of the binding curves of TROY to the antibody FAB. By comparing the binding curves with reaction kinetics models of the antigen-antibody complex association and dissociation rate constants can be extracted as successfully demonstrated by Duan et al.[32] We start with a two-compartment reaction model[33]

$$V\frac{dC_S}{dt} = Vk_m(C_{TROY} - C_S) - S(k_1 C_S(c_{max} - c_{bound}) + k_{-1} c_{bound}) \quad (2a)$$

$$\frac{dc_{bound}}{dt} = k_1 C_S(c_{max} - c_{bound}) - k_{-1} c_{bound} \quad (2b)$$

where $V$ is the compartment volume, $S$ is the surface area, $C_S$ is the antigen concentration in the surface compartment, $c_{max}$ is the concentration of antibody FABs on the surface and $c_{bound}$ is the concentration of bound antigen-antibody complex; $k_m$, $k_1$ and $k_{-1}$ are the transport rate constant, the association and the dissociation rate constant, respectively. We have to pay attention to the fact that kinetics are both evaluated in the volume and on the surface, hence $C_S$ and $C_{TROY}$ have dimensions of a volume concentration and $c_{max}$ and $c_{bound}$ have dimensions of a surface concentration. No analytical solutions are known to these equations, but, under certain assumptions, we can simplify them and fit the binding data to the model. In our case a fast-mixing model for adsorption can be applied which is not diffusion-limited. In the following we will justify this choice.

A measure for the ratio of convective transport rate to diffusive transport rate is the dimensionless Peclét number. There are two Peclét numbers associated with our experiment,[34] one for our fluid chamber design ($Pe_c$) and one that takes the nanowire dimensions into account ($Pe_n$). They depend on the volumetric flow rate $Q$, the chamber width $w_c$ and height $h$, the diffusion constant of the antigen $D$ and the nanowire width $w_n$. The two Peclét numbers are then calculated[34]

$$Pe_c = \frac{Q}{w_C D}, \quad (3a)$$

$$Pe_n = 6\left(\frac{w_n}{h}\right)^2 Pe_c. \quad (3b)$$

The parameters of our setup are $Q$ = 30 µl/min, $w_c$ = 1 mm, $h$ = 0.1 mm, $w_n$ = 20 × 300 nm, since we have a set of 20 wires. Assuming a diffusion constant of 10 µm²/s we obtain a Peclét number for the chamber $Pe_c$ = 50,000 >> 1. This indicates that antigen transport in the chamber is dominated by convection. The second Peclét number $Pe_n$ = 900 is still much larger than 1 and hence convection is also the dominant transport mechanism to the sensor and supply of antigen to the nanowire is not limited by diffusion. This means that the concentration in the surface compartment is essentially the same as the volume concentration and equation (2) simplifies to:

$$\frac{dc_{bound}}{dt} = k_1 C_{TROY}(c_{max} - c_{bound}) - k_{-1} c_{bound}. \quad (4)$$

This can be solved analytically and then the conductance change $\Delta G$ is proportional to

$$\Delta G \propto 1 - e^{-(k_1 C_{TROY} + k_{-1})t}. \quad (5)$$

For desorption the conductance change is proportional to

$$\Delta G \propto e^{-k_{-1} t} \quad (6)$$

Equation (5) was used to approximate the real-time antigen-antibody binding data (Figure 3a). The curves were time-shifted to ensure that injection of the buffer containing TROY happens at time zero. The decay constants $\tau_{adsorption}$ and $\tau_{desorption}$ were then plotted for the different concentrations to extract the association and dissociation constants (Figures 3b and 3c). Due to the small value of $k_{-1}$ its calculation from the binding curve is not very accurate. It is thus calculated from the unbinding curve using equation (6). The association rate constant is (1 ± 0.2) × 10⁵ mls⁻¹g⁻¹) or (5 ± 1) × 10⁶ s⁻¹M⁻¹. The dissociation rate constant is 0.13 ± 0.01 s⁻¹. From those two values, the equilibrium constant can be calculated according to $K_{eq} = k_{-1} \times k_1^{-1}$ = 26 nM ± 7 nM.

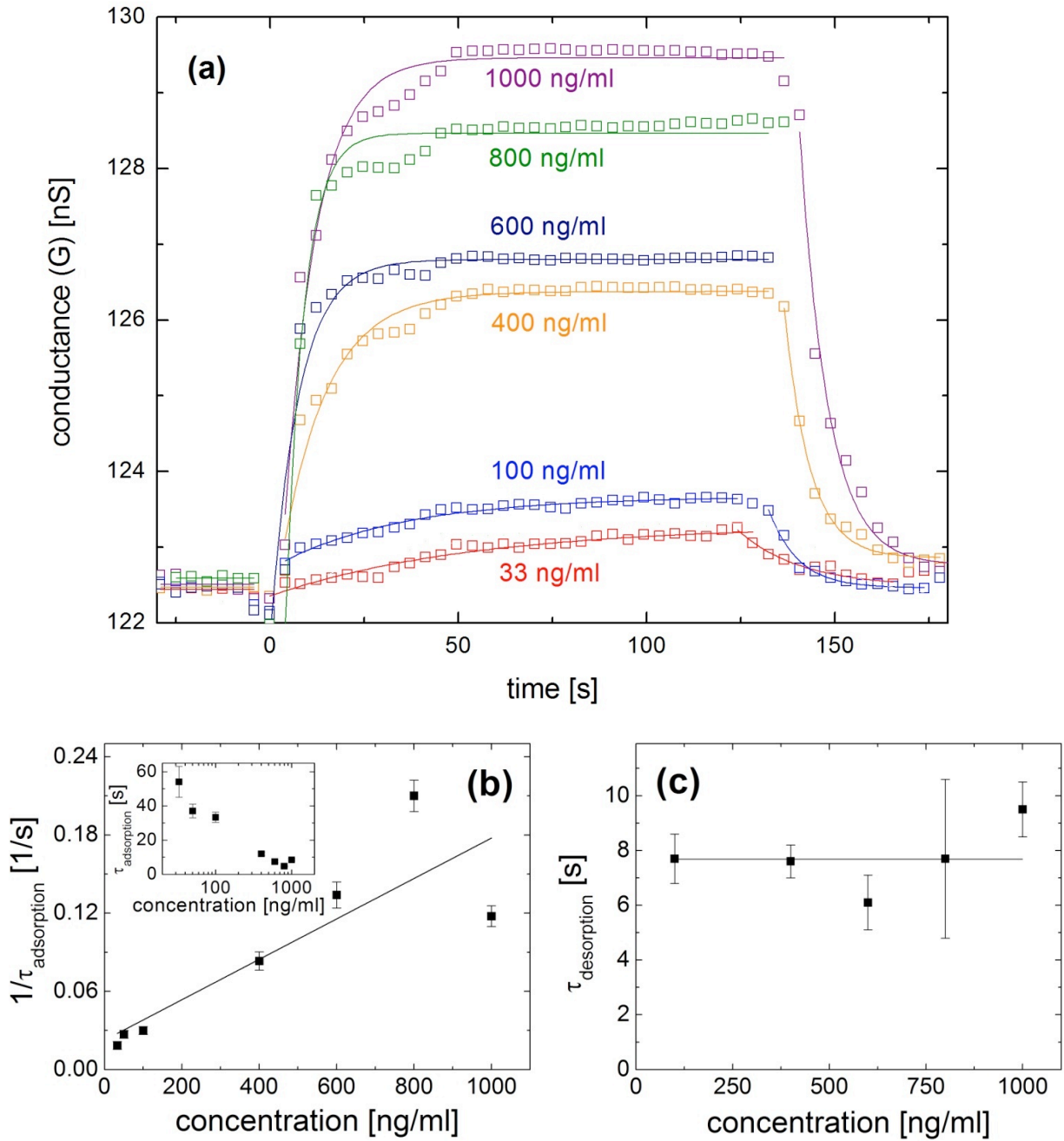

**Figure 3.** (a) Comparison of the nanowire response curves to different concentrations of TROY. The graphs were aligned to show injection of buffer containing TROY at the same time (0 s) and starting with the same baseline. The errors are estimated from the signal drift to 0.3 nS. For visual clarity they are not shown in the graph. The data is approximated by a linear fit for the baseline and by a fast mixing reaction kinetics model (equations (5) and (6)) for binding and unbinding. (b) The inverse time constants of the exponential approximations to the adsorption data in (a) for different concentrations. The association and dissociation rate constants can be extracted from the linear fit according to $1/\tau_{adsorption} = k_1 C_{TROY} + k_{-1}$. The inset shows the time constants $\tau_{adsorption}$ for different concentrations. (c) The time constants for the desorption data in (a). Error bars in (b) and (c) are statistical errors from the fit.

For comparison with the reaction kinetics of TROY binding to the antibody FAB, the binding curves for BSA are shown in Figure 4a and approximated by exponential decays. Because the binding is nonspecific the time constants are much bigger compared to binding of TROY especially for concentrations greater than 1 μg/ml. This is emphasized in Figure 4b. To enable direct comparison, the conductance signals were normalized by subtraction of the saturation conductance $G_{sat}$ and subsequent division by the total conductance change $\Delta G_{total}$. This normalized signal $G_{norm}$ was plotted logarithmically to visualize the different time constants. The difference of the binding kinetics for high concentrations of BSA and TROY enables a distinction between specific and non-specific binding.

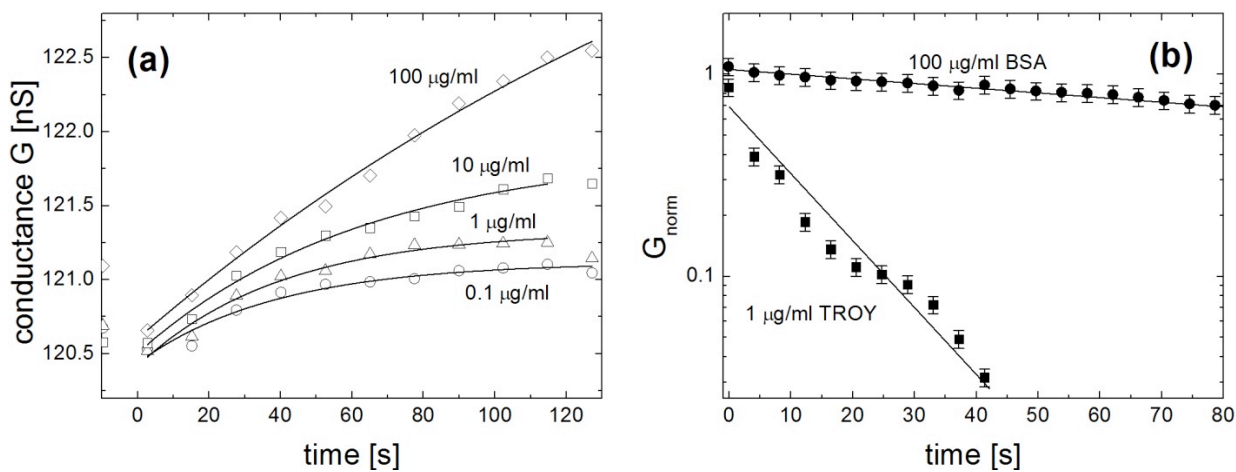

**Figure 4.** Comparison of the adsorption of TROY and BSA. (a) The nanowire conductance change for different concentrations of BSA. The graphs were aligned to show injection of buffer containing BSA at the same time (0 s) and starting with the same baseline. The data is binned (3 adjacent original data points are averaged and summarized into 1). The lines are exponential fits. (b) The data is normalized by the difference between baseline and saturation value for the conductance and displayed logarithmically for easy comparison of the time constants. The errors are estimated from the uncertainty in the saturation and baseline conductance. The lines are guides to the eye. It is obvious that the time constants are different which enables a distinction between specific and nonspecific binding.

**IV. Conclusion**

In conclusion, we have shown the detection of TROY in buffer solution using antibody-functionalized silicon nanowires. TROY is expressed in primary and metastatic melanoma cells and shed into the surrounding environment. It may thus serve as a biomarker for melanoma. The detection limit was about 200 pM. We have demonstrated specificity using BSA as control. For similar concentrations as TROY, up to 1 μg/ml, the signal due to BSA was below the detection limit. For larger concentrations of BSA, a distinction to TROY is possible by comparing the reaction kinetics. The adsorption of TROY, which could be modeled by a fast-mixing two-compartment reaction model, exhibits much smaller time constants than an exponential fit for the BSA adsorption. The association and dissociation rate constants for TROY-antibody binding on the nanowire surface were calculated to $(5 \pm 1) \times 10^6 \text{ s}^{-1}\text{M}^{-1}$ and $0.13 \pm 0.01 \text{ s}^{-1}$, respectively. For sensing in solutions containing high concentrations of salt antibody fragments were used. In further studies this hypothesis should be confirmed and eventually TROY should be detected in serum and validated as a melanoma biomarker to advance the technique towards a POC device that would be expected to revolutionize melanoma patient care.

AUTHOR INFORMATION
**Corresponding Author**
*(P. Mohanty at mohanty@physics.bu.edu)

**Acknowledgement**: The authors thank Biosite Diagnostics (San Diego, CA) for providing TROY antibodies. The authors acknowledge NIH, NSF and Batelle Memorial Institute for the financial support of this work.